# A COMPARATIVE ANALYSIS OF THERMAL FLOW SENSING IN BIOMEDICAL APPLICATIONS

Baseerat Khan, Suhaib Ahmed and Vipan Kakkar

Department of Electronics and Communication Engineering, Shri Mata Vaishno Devi University, Katra, India

## ABSTRACT

*Flow sensors have diverse applications in the field of biomedical engineering and also in industries. Micromachining of flow sensors has accomplished a new goal when it comes to miniaturization. Due to the scaling in dimensions, power consumption, mass cost, sensitivity and integration with other modules such as wireless telemetry has improvised to a great extent. Thermal flow sensors find wide applications in biomedical such as in hydrocephalus shunts and drug delivery systems. Infrared thermal sensing is used for preclinical diagnosis of breast cancer, for identifying various neurological disorders and for monitoring various muscular movements. In this paper, various modes of thermal flow sensing and transduction methods with respect to different biomedical applications are discussed. Thermal flow sensing is given prime focus because of the simplicity in the design. Finally, a comparison of flow sensing technologies is also presented.*

## 1. INTRODUCTION

Flow sensors have a great utility in industrial, biomedical and research fields for monitoring and control operations. Flow sensors that are based on micromachining add aflavor to technologies such as valves, fluidic channels, heater elements for creating a complete micro analysis system e.g.; implantable micro-pumps can be used to deliver drugs without the need for inserting needle in human skin[2].

Fluid flow can be measured by different methods and can be categorized at macro-scale and micro-scale level. Mechanical flow-meters such as venture-meter, positive displacement meters and orifice are used to measure flow rate at macro levels and are extensively used for industrial purpose. Because of their moving parts, interference with the fluid flow, and difficulty in fabrication, these flow sensors can't be used for micro scale applications such as in biomedical.
At micro level, micromachined flow sensors are without any moving parts and hence are easy for fabrication process and also simplify the operational requirements. As a result, great interest is being developed in the MEMS community for fabrication of micro sensors. Micromachined flow sensors are used to check the oscillation of the fluid in fluidic micro-channels [1].This allows monitoring the flow behaviour in real time and avoiding misdiagnosis. The flow sensors can also be categorized as electronic sensors based on operation principles. This includes thermal and ultrasonic sensors.

Thermal flow sensors measure the flow rate of a fluid by measuring the changes in thermal phenomenon by heat transfer such as conduction, convection or radiation. The signal can be transduced into the electrical signal (voltage) to measure the sensor response to the flow. The flow sensors that are of micro range are gaining prime importance because of lower power consumption, high sensitivity to flow rates and can easily be operated in different modes. Moreover, these sensors can measure the thermal properties of fluid (e.g. blood or CSF) such as conductivity or diffusivity [3].

Thermal flow sensors are thermally isolated so that heat loss only due to convective cooling should occur. Losses due to substrate and other metallic contacts degrade the sensor performance. The sensitivity and power consumption can be optimized by varying the shape and position of the sensor. Also, a suitable material should be chosen for the design of flow sensor.

Ultrasonic sensors measure the flow such as blood flow by using the propagating principle of ultrasound signal in tissue. The ultrasound frequencies are in the range of above 20 kHz. These sensors find wide applications in medicine such as in imaging. Doppler sensors and time of flight sensors are sub-categories of ultrasonic sensors. Doppler sensors measure the flow velocity by measuring the shift in frequency, commonly known as Doppler shift between the transmitted and received signal. The equation for determining velocity is:

$$V_d = \frac{cf_d}{2\cos\theta f_0}$$

Where,
$f_d$ is the Doppler frequency,

$f_o$ is incident frequency, and $\theta$ is the angle of placement of sensor with respect to vessel.

Time of flight sensors measure the flow by measuring the difference between the transit times. An implantable time of flight sensor using electrochemical impedance was developed for in-vivo monitoring within shunts for hydrocephalic patients [5].

## 2. MODES OF THERMAL SENSING

Various modes of flow sensing are: Calorimetric, anemometric and thermal time of flight are the three modes of flow sensing:

### 2.1. Anemometric sensors

It operates by transferring heat from a locally heated element such as micro-heater to the fluid flowing around it. Hot-film and hot-wire sensors operate using the same physical principle. Hot film uses a resistive thin film having different values of resistance depending upon the material used whereas, hot wire uses simply a metal wire. The rate at which convective heat loss occurs can be measured and correlated to flow velocity. Thermal properties, heater temperature and velocity of the fluid dictates the amount of heat transferred to the fluid. Figure 1 gives an illustration of anemometric sensing.

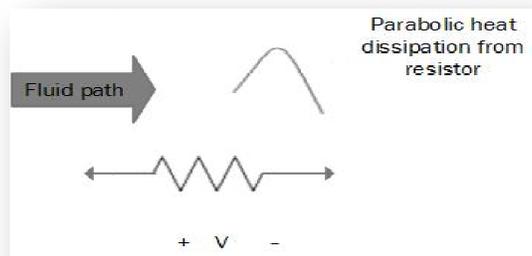

Figure 1. Anemometric Sensor

The material chosen for heating element should have high thermal coefficient of resistance, as sensitivity of the sensor is proportional to material's coefficient of resistance. Nichrome, platinum are the preferable choices as they are biocompatible and are more commonly used in biomedical applications such as in medical tubing and implantable devices. Also, platinum is resistant to corrosion and compatible with micromachining techniques.

Anemometric sensors operate in either constant temperature mode or constant power mode. In constant temperature mode, the temperature is held constant by controlling the power dissipated by the sensor. With the increase in fluid flow rate, the thermal conductance of the element increases and more heat is dissipated. However, in constant power mode, the power remains constant by applying a constant current bias to the heating element and changes in resistance or voltage are correlated to the flow rate.Constant temperature mode is preferred over constant power because of the better sensitivity and frequency response [3].

### 2.2. Calorimetric flow sensor

Calorimetric sensing requires a thermal sensor placed upstream and downstream to the heating element. The variation in upstream and downstream temperature due to the flow gives an indirect measure of fluid velocity. The working of a calorimetric sensor is illustrated in figure 2.

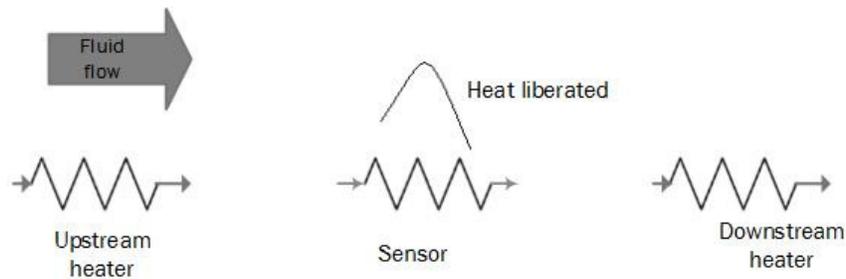

Figure 2. Calorimetric Sensor [17]

Greater the flow rate, greater will be the variation between the upstream and downstream temperatures.

### 3.3. Thermal time of flight

In this mode of sensing, a thermal pulse is generated by a hot wire and injected into the fluid stream. The downstream sensor detects the heat pulse which is carried away from source by the fluid flow. The time it takes for an input signal from generation to detection (transit time) can be used to determine the flow rate. Thermal conductivity, flow velocity and distance between heating element and sensing element are certain factors which have a great influence on transit time. Figure 3 gives an insight into working of time of flight sensor.

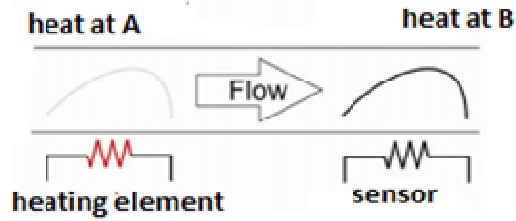

Figure 3. Thermal Time of Flight [17]

## 3. DIFFERENT TRANSDUCTION PRINCIPLES

Thermal flow sensors use thermoelectric, thermoresistive or acoustic principles for transducing mechanism.Thermoresistive sensors simply use a resistor to which a certain potential is applied at the terminals. When the fluid flows over the sensor, convective heat loss occurs and as such there is a drop in temperature. Decrease in temperature causes a substantial decrease in resistance, which causes change in voltage or current. This change can be calibrated to the fluid flow. These sensors are by far used widely because of its easy fabrication and operability.

Thermoelectric sensors use a heating element in conjunction with the thermocouples. When thermocouples are connected in series, they form thermopiles. The output voltage of a thermopile due to temperature changes is the summation of individual output voltages of series connected thermocouple. The output voltage is determined from the equation given below:

$$V_{ab} = (\alpha_a - \alpha_b)(T_a - T_b)$$

Where,

$\alpha_a$ is seebeck coefficient of material 'a';
$\alpha_b$ is seebeck coefficient of material 'b';
$T_a$ is hot junction temperature, and
$T_b$ is the cold junction temperature.

However, these are not much used because of the less conventional materials required for its fabrication, which makes it complicated. Also with the increase in number of thermocouples, the performance of sensor degrades because of the increase in Johnson noise. Polysilicon and metals are the commonly used materials for thermoelectric transduction.

Surface acoustic wave flow sensors and cantilevers sense the changes in the resonant frequency of the substrate in response to the stresses applied to the material. A novel wave flow sensor was developed using a quartz material. ST-X quartz material was chosen because of excellent temperature stability. The test results showed a change of 4.24 degree/kpa in terms of pressure and a phase change of 1 degree for 11.8 ml/min change in flow rate [6].From fabrication point of view SAW sensors use conventionalpiezoelectric materials such as ST-X quartz. Low power consumption, easy operability and compatibility with wireless module are certain features of SAW sensors. However, it has limited compatibility with the CMOS devices and also poses certain potential constraints.

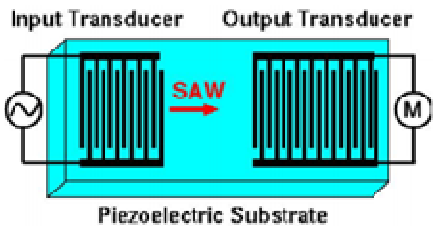

Figure 4. SAW port structure [6]

## 4. APPLICATIONS OF MICRO THERMAL MACHINED SENSORS

Mailly et al. [7] fabricated two platinum films which were developed as resistors on a substrate .The resistors act as a temperature sensor and a heating element, respectively, to measure the fluid rate. A silicon nitride layer is used as thermal isolation layer over the substrate [7].The thermal and electrical analysis of the model was done.A flexible sensor was mounted on a substrate of parylene, which is a cross linked polymer. The sensor was placed in a conductive catheter and used to measure the pulsatile blood flow in arteries of rabbits [8].

Bailey et al. [9] fabricated a wire made up of platinum having dimensions 100 nm× 2µm×60µm.The frequency and spatial results were better in comparison to traditional anemometers.Later on, carbon nanotubes were added to increase the sensitivity of a sensor.Neto et al. [10] used finite element method software for developing a calorimetric flow meter. The flow was measured by calculating the difference between the upstream and downstream temperature around the heater element. The properties of the heater element were set from the material database. The results were that greater the fluid velocity, greater will be the heat distribution around the heater element. For bio-sensing, the height of the heater element is an important design variable and it should have less interference with the fluid profile [10] the results were then compared with the actual experimental setup.

A flow sensor incorporated in a catheter was presented for real time cerebral blood flow monitoring in certain region of interest in [11]. The sensor is operated in constant temperature mode and employs four wire configurations for periodic heating and cooling mechanism. This approach gives reliable results and ensures zero drift with MEMS based sensors. The results show that the sensor has a sensitivity of 0.973 mV/ml/100 g/min in the range from 0-160 ml/100 g/min [11].Further scope is multiple sensors can be incorporated in a catheter for neuro-modal monitoring.

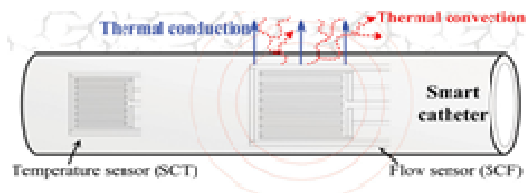

Fig. 5: Smart Catheter flow sensor [11]

A novel flow sensor system was modeled using thermal time of flight principle. The system is regarded as a linear time invariant system, comprising of temperature sensors and a hot wire. Fluid velocity, fluid media and certain heat effects were measured using signal analysis of rise and fall times and also by the thermo fluid measurement. Ecin et al. actually presented the mathematical model with thermodynamic and fluid mechanical properties [12].

In [13], a parylene based micro flow sensor was designed and fabricated using electrochemical impedance transduction technique. Three pair of electrodes were used, one for bubble generation and other two for measurements. This sensor was used for chronic in-vivo monitoring of fluid within shunts for hydrocephalic patients. High precision measurement of micro-bubble, which is used as a tracer, was obtained over the velocity of 0.83-83µm/sec. The results show that the flow rate was inversely proportional to time of flight.

A linearization bridge technique of p-n junction anemometric sensor has been proposed in [14] as the flow characteristic is non-linear. The system has been developed and fabricated and the experimental results show linearity in flow characteristics. In this paper, an electric equivalent of the sensor has been designed.

A MEMS based flexible flow sensor for online monitoring of body fluid such as blood was proposed and integrated with catheter. A resistive heater element of varying geometries was placed on a substrate of different materials and the velocity profile, temperature distribution of fluid around the heater element was analyzed [15].

The development and fabrication of a thermal flow sensor, used for biological and chemical applications has been discussed in [16]. Finite element software was used for the simulation and the simulation showed a sensitivity of µv/µL/min with a resolution of 10µL/min. For an input current of 40mA and heater width of 50µm, the temperature of the fluid increased by 5°C. The results show that the sensor has suitable sensitivity with low interference in fluid flow.

## 5. PERFORMANCE COMPARISON OF THERMAL FLOW SENSORS

The advantages and the shortcomings/challenges of the various thermal flow sensors can be summarized by Table 1.

Table 1. Comparison of different thermal flow sensors

| Sensor Type | Advantages | Challenges |
|---|---|---|
| Calorimetric | • Inherently sensitive to mass flow | • Need for localized heat |
| Anemometric | • Mass flow is sensitive to heat transfer<br>• Simple technique to use<br>• Better performance over a wide range of flow | • Non-linearity in flow |
| Ultrasonic | • No-invasive in nature<br>• Can be used as a clamp on transducer | • Requires precise and stable alignment of beam on blood vessel<br>• Slight deviations or miscalculations can cause erroneous results<br>• Traces or particles are required to reflect the signal |

The most important design parameter is that the substrate should have minimal thermal conductivity so that large amount of heat is not dissipated to it. Polymers have minimal thermal conductivity and are mostly used as substrate material and are also used for packaging. The geometry of the heating element is also a very critical parameter as the sensitivity of a device depends on the conductance, which in turn depends on the size and shape of the sensing element.

## 6. CONCLUSION

Micromachining of thermal flow sensors provides us with a platform for establishing new methodologies. Different modes of transduction and diversity in materials can be used for wide application range. Thermal sensing can be used for diagnosis and monitoring of fluid rates in human body for timely intervention. A simple system controlling current through a resistive element (hot-film or hot-wire) is preferential over other sensor systems. The output parameter such as voltage can be correlated to the fluid velocity. Minimal power consumption, easy operation and fabrication are some of the advantages that make thermal sensing an ultimate choice.

International Journal of Biomedical Engineering and Science (IJBES), Vol. 3, No. 3, July 2016## REFERENCES

[1] N.T Nguyen. "Micro machined flow sensors-a review." *Flow Measurementand Instrumentation*, vol. 8,no.1, pp.7 – 16, 1997.

[2] Waldron, Matthew J., "Design, simulation, and fabrication of a flow sensor for an implantable micro pump" (2009).Thesis, Rochester Institute of Technology.

[3] W.-KMeng, E.F.-C. "MEMS Technology and Devices for a Micro fluid Dosing System". Ph.D. Dissertation, California Institute of Technology, Pasadena, CA, USA, 2003, pp. 150.

[4] John G. Webster, "Medical Instrumentation-Application and Design", John Wiley and Sons, Singapore (2005).

[5] Yu, L.; Kim, B.J.; Meng, E., "An implantable time of flight flow sensor," in *Micro Electro Mechanical Systems (MEMS), 2015 28th IEEE International Conference on*, pp.620-623, 18-22 Jan. 2015.

[6] Yizhong Wang; Zheng Li; Lifeng Qin; Chyu, M.K.; Qing-Ming Wang, "Surface acoustic wave flow sensor," in *Frequency Control and the European Frequency and Time Forum (FCS), 2011 Joint Conference of the IEEE International* , pp.1-4, 2-5 May 2011.

[7] Mailly, F., et al., "Anemometer with hot platinum thin film," *Sensors and Actuators, A: Physical,* vol.94,no.(1-2), pp.32-38,2001.

[8] Bailey, Sean CC, et al. "Turbulence measurements using a nanoscale thermal anemometry probe." *Journal of Fluid Mechanics* 663 (2010): 160-179.

[9] Wu, S.; Lin, Q.; Yuen, Y.; Tai, Y.-C. "MEMS flow sensors for Nano-fluidic applications". *Sens. Actuat. A .*vol*. 89*, pp.152–158,2001.

[10] Barreto Neto, A.G.S.; Lima, A.M.N.; Moreira, C.S.; Neff, H., "Design and theoretical analysis of a bidirectional calorimetric flow sensor," in *Instrumentation and Measurement Technology Conference (I2MTC) Proceedings, 2014 IEEE International*,pp.542-545, 12-15 May 2014.

[11] Chunyan Li; Pei-Ming Wu; Zhizhen Wu; Ahn, C.H.; Hartings, J.A.; Narayan, R.K., "Smart catheter flow sensor for continuous regional cerebral blood flow monitoring," in *Sensors, 2011 IEEE*, pp.1417-1420, 28-31 Oct. 2011.

[12] Ecin, O.; Zhao, R.; Hosticka, B.J.; Grabmaier, A., "Thermo-fluid dynamic Time-of-Flight flow sensor system," in *Sensors, 2012 IEEE*, Taipei,pp.1-4, 28-31 Oct. 2012.

[13] Yu, L.; Kim, B.J.; Meng, E., "An implantable time of flight flow sensor," in *Micro Electro Mechanical Systems (MEMS), 2015 28th IEEE International Conference on*, Estoril,2015,pp.620-623, 18-22 Jan. 2015.

[14] Bera, S.C.; Marick, S., "Study of a Simple Linearization Technique of p-n-Junction-Type Anemometric Flow Sensor," in *Instrumentation and Measurement, IEEE Transactions on* , vol.61, no.9, pp.2545-2552, Sept. 2012.

[15] Maji, D.; Das, S., "Simulation and feasibility study of flow sensor on flexible polymer for healthcare application," in *Point-of-Care Healthcare Technologies (PHT), 2013 IEEE*, pp.132-135, 16-18 Jan. 2013.

[16] Mielli, M.Z.; Carreno, M.N.P., "Thermal flow sensor integrated to PDMS-based microfluidic systems," in *Microelectronics Technology and Devices (SBMicro), 2013 Symposium on*, pp.1-4, 2-6 Sept. 2013.

[17] Jonathan T.W.Kuo,Lawrence Yu,Ellis Meng, "Micromachined thermal flow sensors-A Review",*Micromachines,*vol. 3*,*no.3*,* pp. *570-573,2012.*

[18] Wu, S.; Lin, Q.; Yuen, Y.; Tai, Y.-C. "MEMS flow sensors for nano-fluidic applications", *Sens. Actuat.,* vol.89, pp. 152–158,2001.

[19] Yu, H.; Ai, L.; Rouhanizadeh, M.; Patel, D.; Kim, E.S.; Hsiai, T.K. "Flexible polymer sensors for *in vivo* intravascular shear stress analysis" *J. Microelectromechanical Syst.*,vol.17,pp.1178–1186,2008.

[20] Meng, E.; Li, P.-Y.; Tai, Y.-C, "A biocompatible Parylene thermal flow sensing array," *Sens. Actuat. A*,vol. *144*,pp. 18–28,2008
7